%% file: main.tex
\documentclass[10pt,conference]{IEEEtran}
\usepackage[T1]{fontenc}
\usepackage[utf8]{inputenc}
\usepackage{microtype}
\usepackage{amsmath,amssymb,amsfonts}
\usepackage{graphicx}
\usepackage{xcolor}
\usepackage{cite}
\usepackage{url}
\usepackage{balance}     
\usepackage[hidelinks]{hyperref}    

\usepackage{algorithm}
\usepackage{algpseudocode}
\usepackage{enumitem}
\setlist[itemize]{leftmargin=*}
\setlist[enumerate]{leftmargin=*}
\usepackage{tikz}
\usetikzlibrary{calc,arrows.meta,positioning,fit,shapes.geometric}
\usepackage{listings}
\usepackage{xspace}
\usepackage{booktabs}
\usepackage{multirow}
\usepackage{tabularx}
\usepackage{ragged2e}
\usepackage{array}
\usepackage{siunitx}
\usepackage{threeparttable}
\newcolumntype{Y}{>{\RaggedRight\arraybackslash}X}

\newcommand{\analysisagent}{Analysis\-Agent\xspace}
\newcommand{\benchmark}{Analysis\-Bench\xspace}

\newcommand{\mypara}[1]{\vspace{0.1in}\noindent\textbf{#1}}

\input{results-macros}

\lstdefinestyle{terminal}{
	basicstyle=\ttfamily\footnotesize,
	breaklines=true,
	frame=tb,
	rulecolor=\color{gray!60},
	backgroundcolor=\color{gray!6},
	xleftmargin=0.5em,
	framexleftmargin=0.5em,
	aboveskip=0.8\medskipamount,
	belowskip=0.8\medskipamount,
	language=bash,
	commentstyle=\color{gray!70}\itshape,
	keywordstyle={},
	columns=flexible,
	keepspaces=true
}

\lstdefinestyle{dockerfile}{
	basicstyle=\ttfamily\footnotesize,
	breaklines=true,
	frame=tb,
	rulecolor=\color{gray!60},
	backgroundcolor=\color{blue!4},
	xleftmargin=0.5em,
	framexleftmargin=0.5em,
	aboveskip=0.8\medskipamount,
	belowskip=0.8\medskipamount,
	language=bash,
	morekeywords={FROM,RUN,WORKDIR,ENV,COPY,CMD,SHELL,USER},
	keywordstyle=\color{blue!70!black}\bfseries,
	commentstyle=\color{gray!70}\itshape,
	columns=flexible,
	keepspaces=true
}

\lstdefinestyle{json}{
	basicstyle=\ttfamily\footnotesize,
	breaklines=true,
	frame=tb,
	rulecolor=\color{gray!60},
	backgroundcolor=\color{gray!6},
	xleftmargin=0.5em,
	framexleftmargin=0.5em,
	aboveskip=0.8\medskipamount,
	belowskip=0.8\medskipamount,
	columns=flexible,
	keepspaces=true,
	string=[s]{"}{"},
	stringstyle=\color{orange!80!black},
	comment=[l]{:},
	commentstyle=\color{black},
}

\lstdefinestyle{plaintext}{
	basicstyle=\ttfamily\footnotesize,
	breaklines=true,
	frame=tb,
	rulecolor=\color{gray!60},
	backgroundcolor=\color{gray!6},
	xleftmargin=0.5em,
	framexleftmargin=0.5em,
	aboveskip=0.8\medskipamount,
	belowskip=0.8\medskipamount,
	columns=flexible,
	keepspaces=true
}

\title{Evaluating LLM Agents on\\ Automated Software Analysis Tasks}

\author{
\IEEEauthorblockN{Islem Bouzenia}
\IEEEauthorblockA{\textit{CISPA Helmholtz Center for Information Security}\\
Germany\\
bouzenia.islem@pm.me}
\and
\IEEEauthorblockN{Cristian Cadar}
\IEEEauthorblockA{\textit{Imperial College London}\\
United Kingdom\\
c.cadar@imperial.ac.uk}
\and
\IEEEauthorblockN{Michael Pradel}
\IEEEauthorblockA{\textit{CISPA Helmholtz Center for Information Security}\\
Germany\\
michael@binaervarianz.de}
}

\begin{document}

\maketitle

\begin{abstract}
Numerous software analysis tools exist today, yet applying them to diverse open-source projects remains challenging due to environment setup, dependency resolution, and analysis tool configuration.
LLM-based agents offer a potential solution, yet their effectiveness on the specific task of \emph{automated software analysis} has not been systematically studied.
Unlike issue solving or general environment setup, this task requires installing and configuring a separate analysis tool alongside the target project, generating tool-specific prerequisites, and validating that the analysis tool produces meaningful analysis outputs rather than merely declaring successful termination without evidence.
We introduce \benchmark, a benchmark of \benchTotalTasks{} tool-project pairs spanning \benchNumTools{} analysis tools and \benchNumProjects{} diverse C/C++ and Java projects, each with a manually constructed reference setup.
Using \benchmark, we evaluate four agent architectures across four LLM backends.
Our custom agent, \analysisagent, achieves \emph{manually verified} success rates of \bestVerifiedRate\% (Gemini-3-Flash, \bestVerifiedCount/\benchTotalTasks{} tasks), compared to \resExecGeminiVerified\% for the best baseline (ExecutionAgent).
Beyond quantitative results, we identify key limitations in existing agents, including stage mixing, poor error localization, and premature termination, and show that agentic architecture plays a critical role beyond LLM capability alone.
We further find that whole-program analyses and Java-specific tools are the most difficult tasks, that Java toolchains pose greater challenges than C/C++, and that self-validated success consistently overstates manually verified success.
Extended runs with \analysisagent-produced setups also surface two previously unknown defects in masscan and radare2, confirming that the produced setups support real downstream analysis.
\end{abstract}

\section{Introduction}

To support developers in ensuring software quality, security, and performance, decades of research and engineering have produced a wide variety of \emph{software analysis tools}.
For example, developers can choose from a diverse range of static analyzers~\cite{Vallee-Rai1999,ErrorProne,Banerjee2019}, symbolic execution engines~\cite{Sen2005,Cadar2008,Stephens2016}, fuzzers~\cite{afl2013,fioraldi2020afl++}, and profilers~\cite{Graham1982,Xu2009,Curtsinger2015,icpe2021}.
When adopted widely, these analysis tools have a proven impact. For instance, Google's static analysis platform Tricorder surfaces hundreds of thousands of warnings per day across the company's codebase~\cite{Sadowski2018}, and OSS-Fuzz has found thousands of vulnerabilities in open-source projects through continuous fuzzing~\cite{OSSFuzz}.

Unfortunately, applying a given analysis tool to a new project is notoriously difficult~\cite{Bessey2010}: the analysis tool requires a compatible environment with specific compilers, libraries, and configuration, while the target project often needs special build flags or tool-specific prerequisites such as LLVM bitcode or a running JVM. Both must be co-located in an isolated, reproducible container, and developers must still verify that the analysis ran on the target code and produced meaningful outputs, since a misconfigured setup may process only trivial inputs and give a false sense of success. Prior work has documented this setup burden for static analyzers~\cite{Bessey2010,johnson2013don}, fuzzers~\cite{fioraldi2020afl++}, and symbolic execution engines~\cite{Cadar2008}, contributing to limited adoption in practice.
LLM-based agents, with their ability to plan and execute shell commands, interpret error messages, and adapt to unforeseen failures, offer a promising approach to this problem.
Agents have already demonstrated success in related tasks, including issue solving~\cite{Hu2024,Yang2024a,Zhang2024a,Wang2024a}, automated program repair~\cite{icse2025-RepairAgent,Rondon2025,Cheng2025}, web automation~\cite{Wang2024a}, and environment construction~\cite{issta2025_ExecutionAgent,Milliken2025,Eliseeva2025,Bogin2024}.
However, no prior work has studied the end-to-end task of automated software analysis in a controlled, multi-tool setting.
Unlike project-only environment setup~\cite{issta2025_ExecutionAgent}, automated software analysis requires installing and configuring a separate analysis tool alongside the target project, producing tool-specific prerequisites (e.g., LLVM bitcode, classpath JARs), and validating that the analysis tool produces meaningful outputs rather than merely running without errors.
As we show in our evaluation, changing the task description of an existing agent is not sufficient: the best baseline averages \resExecAvgVerified\% verified success across LLM backends, compared to \resAAAvgVerified\% for \analysisagent.

To study LLM-based agents on this task, we construct \emph{\benchmark{}}, the first benchmark for end-to-end \emph{automated software analysis}.
\benchmark evaluates whether an agent can set up and execute diverse analysis tools on open-source projects from scratch, spanning the full pipeline from environment construction to evidence validation, where success requires concrete analysis artifacts rather than only a zero exit code.
The benchmark consists of \benchTotalTasks{} tool-project pairs spanning \benchNumTools{} analysis tools and \benchNumProjects{} diverse C/C++ and Java projects (Tables~\ref{tab:tools} and \ref{tab:repo-characterizations}).
Each task includes a manually verified reference setup and success criteria for rigorous evaluation.

We compare three baselines, RAG-Agent~\cite{lewis2020retrieval}, Mini-SWE-Agent~\cite{Yang2024a}, and ExecutionAgent~\cite{issta2025_ExecutionAgent}, across four LLMs (GPT-5-nano, GPT-5-mini, DeepSeek-V3.2, Gemini-3-Flash).
This comparison reveals both promise and systematic limitations in existing agents: \emph{stage mixing} (interleaving unrelated workflow steps, e.g., attempting analysis before the analysis tool is installed), \emph{poor error localization} (verbose logs obscuring root causes), and \emph{premature termination} (stopping after partial evidence of success without validating the final analysis output; \S\ref{subsec:baseline_limitations}), with the best baseline (ExecutionAgent + Gemini-3-Flash) reaching \resExecGeminiVerified\% verified success.

To address these limitations, we present \emph{\analysisagent{}}, a novel agent designed specifically for the automated software analysis task.
Our approach incorporates three design principles that each address one of the above limitations:
explicitly staged execution, single-action cycles with log condensation, and evidence-based validation.
Empirical evaluation across four LLM backends shows that \analysisagent with Gemini-3-Flash achieves the highest verified success rate across all configurations (\bestVerifiedRate\%, \bestVerifiedCount/\benchTotalTasks{} tasks).
Because failed runs consume \effFailCycleMult$\times$ more cycles and cost \effFailCostMult$\times$ more than successful runs in our evaluation, the higher success rate of \analysisagent also improves the overall cost profile by reducing failures.

Our contributions are \benchmark, a benchmark of \benchTotalTasks{} tool-project pairs with manually verified reference setups; an empirical study of four agentic architectures across four LLMs; and \analysisagent, a purpose-built agent combining staged execution, log condensation, and evidence-based validation that outperforms all baselines regardless of LLM backend.
We further show that \analysisagent-produced setups support real downstream analysis: extended runs with AFL++ and KLEE surface previously unknown defects in masscan and radare2, respectively.

\section{\benchmark}
\label{sec:task_formalization}

\subsection{Task Definition}
\label{subsec:task_definition}

The \emph{automated software analysis} task requires automatically applying a software analysis tool to a real software project.

\paragraph{\textbf{Inputs}}
Each task is defined by a tool-project pair $(T, P)$ and an execution interface.
\begin{itemize}
\item \emph{Tool specification $T$.} Analysis tool name and acquisition method (e.g., repository URL or a released package).
  
\item \emph{Target project $P$.} Repository URL and pinned revision (commit hash or tag) of the target project to be analyzed.

\item \emph{Execution interface.} A writable workspace, environment interaction tools (e.g., terminal), and a budget (\budgetTimeoutHours, \$\budgetCostCap{} per task) ensuring comparable evaluation.
\end{itemize}

\paragraph{\textbf{Terminology}}
Throughout this paper, we use \emph{artifacts} for concrete build and analysis outputs (e.g., bitcode files, analysis reports), and \emph{evidence} for artifacts that demonstrate meaningful analysis completion on the target project.

\paragraph{\textbf{Task objective}}
Given $(T, P)$, the agent must
(i)~provision an isolated Docker container,
(ii)~install the analysis tool and its dependencies,
(iii)~fetch and build (when required) the target project,
(iv)~execute the analysis tool on the target project, and
(v)~emit evidence that the analysis tool processed project-specific inputs (e.g., source files) and produced tool-specific output artifacts (e.g., warnings) beyond trivial outputs such as analysis tool version strings or help messages.

\paragraph{\textbf{Outputs}}
A successful run produces an evidence package:
\begin{itemize}
	\item \emph{Reproducible environment.} E.g., a Dockerfile.

	\item \emph{Complete execution trace.} The full transcript of executed commands, including stdout/stderr, return codes, and key environment metadata (analysis tool version, OS image, commit hashes).
	
	\item \emph{Tool-specific analysis artifacts.} Logs and output files from executing $T$ on $P$ (e.g., warnings, test cases, call graphs).
\end{itemize}

\paragraph{\textbf{Evidence-based success criteria.}}
We consider a task solved successfully if the environment is reproducible, the analysis tool is installed and runnable, the target project is built and prepared, the analysis tool is invoked on project-relevant inputs, and the outputs contain verifiable, project-specific analysis evidence (e.g., warnings referencing project paths, generated test cases, or call graphs).
This criterion measures \emph{configuration selection}: whether the agent chooses a valid configuration that runs the analysis tool and yields meaningful, project-specific evidence. It does not measure \emph{configuration optimization}, i.e., tuning the tool for maximal analysis quality (e.g., highest coverage or fewest false positives), which depends on tool- and project-specific objectives and is beyond our scope.
We operationalize these criteria through tool-specific validation checks detailed in \S\ref{subsec:manual_ground_truth}.

\subsection{Benchmark Construction}
\label{subsec:benchmark_dataset}

\begin{table}[t]
	\centering
	\caption{Analysis tools in \benchmark{}.}
	\label{tab:tools}
	\small
	\renewcommand{\arraystretch}{1.15}%
	\begin{tabularx}{\columnwidth}{@{} >{\raggedright\arraybackslash}p{1.1cm}
                        >{\raggedright\arraybackslash}p{0.7cm}
                        >{\raggedright\arraybackslash}p{1.3cm}
                        Y @{}}
		\toprule
		\textbf{Tool} & \textbf{Lang.} & \textbf{Technique} & \textbf{Output} \\
		\midrule
		AFL++~\cite{fioraldi2020afl++} & C/C++ & Fuzzing & Coverage data, queued test cases, exercised inputs, hangs/crashes. \\
		KLEE~\cite{Cadar2008} & C/C++ & Symbolic execution & Generated tests (\texttt{.ktest}), path exploration statistics, executed instructions, and coverage. \\
		CSA~\cite{CSA} & C/C++ & Static analysis & Structured reports with source-level warnings and bugs found. \\
		cflow~\cite{cflow} & C & Structural analysis & Text-based call graphs from C sources. \\
		Infer~\cite{facebook-infer} & Java & Static analysis & Issue reports (\texttt{infer-out/}) from whole-program analysis. \\
		WALA~\cite{WALA} & Java & Static analysis & Call graphs and pointer-analysis results from bytecode. \\
		SJK~\cite{SJK} & Java & Profiling & Thread dumps, heap histograms, sampling profiles via JVM attach. \\

		\bottomrule
	\end{tabularx}
\end{table}

\subsubsection{Analysis Tools}
\label{subsubsec:tools}

Table~\ref{tab:tools} summarizes the \benchNumTools{} analysis tools included in \benchmark and the types of outputs they are expected to produce.
We select analysis tools based on three criteria:
(i)~they are open-source and freely available,
(ii)~they represent distinct analysis techniques (fuzzing, symbolic execution, static analysis, structural analysis, and profiling), and
(iii)~they require non-trivial setup beyond a simple package install, which is the core challenge our benchmark targets.
This selection is not exhaustive; important tools such as Valgrind, AddressSanitizer, and SpotBugs are outside the current benchmark.

\subsubsection{Target projects}
\label{subsubsec:projects}

The project rows of Table~\ref{tab:repo-characterizations} show the \benchNumProjects{} selected projects, split between \benchCppProjects{} C/C++ and \benchJavaProjects{} Java codebases.
We select target projects that were actively maintained at the time of collection and cover diverse build systems and dependencies.
To find candidates, we search GitHub for popular repositories using keywords such as ``command-line'', ``program'', and ``cmd'', sorting results by star count; this process is used to identify suitable candidates rather than to construct a statistically representative sample.
We specifically target \emph{command-line programs} that can be invoked directly after installation, and exclude libraries or frameworks used only as dependencies or submodules.
This scoping is deliberate for two reasons:
First, command-line entry points isolate the per-task setup variance we study from harness-construction variance: analyzing a library generally requires writing a driver or fuzzing harness that exercises its API, which in itself is a separate research subproblem with its own challenges and tool-specific conventions~\cite{Cadar2008KLEE,Fioraldi2020AFL,Sherman2025}.
Second, command-line programs still span substantial diversity in build systems, runtimes, and dependencies (Table~\ref{tab:repo-characterizations}), preserving the setup complexity that our benchmark targets. We discuss this threat to generalizability in \S\ref{sec:threats}. Finally, for each candidate, we manually inspect the README and documentation to verify clear command-line invocation examples before inclusion.

\begin{table}
\centering
\caption{Build systems and dependencies in \benchmark.}
\label{tab:repo-characterizations}
\renewcommand{\arraystretch}{1.05}
\setlength{\tabcolsep}{4pt}
\resizebox{\columnwidth}{!}{%
\begin{tabular}{@{} >{\raggedright\arraybackslash}p{2.3cm}@{\hspace{1.5pt}}
                    >{\raggedright\arraybackslash}p{2.1cm}
                    >{\raggedright\arraybackslash}p{3.2cm} @{}}
\toprule
\textbf{Repository} & \textbf{Build toolchain} & \textbf{Key dependencies} \\
\midrule
\multicolumn{3}{@{}l}{\textit{Analysis tools:}} \\
\midrule
AFL++ & Make; gcc/clang & LLVM/clang (opt.) \\
KLEE & CMake & LLVM; STP/Z3; klee-uclibc \\
CSA & CMake + Ninja/Make & LLVM; zlib (opt.) \\
cflow & Autotools + Make & Standard C library \\
Infer & Buck + opam & OCaml; JDK; Clang/LLVM \\
WALA & Gradle/Maven & JDK \\
SJK & Maven & JDK \\
\midrule
\multicolumn{3}{@{}l}{\textit{C/C++ projects:}} \\
\midrule
curl, ImageMagick, fastfetch, masscan, radare2 & Autotools; CMake; Make; Meson & OpenSSL, zlib, nghttp2, libpng, libjpeg, libpcap, capstone, xxhash \\
\midrule
\multicolumn{3}{@{}l}{\textit{Java projects:}} \\
\midrule
Tika, Closure, Saxon-HE, JMH, Checkstyle & Maven; Bazel & JDK; parser libraries \\
\bottomrule
\end{tabular}%
}
\end{table}

\subsubsection{Tool-Project Pairs}
\label{subsubsec:pairing}

To construct the tool-project tasks in \benchmark, we pair each analysis tool with the projects to which it is applicable: \benchCppTools{} C/C++ analysis tools paired with \benchCppProjects{} C/C++ projects ($\benchCppTools \times \benchCppProjects = \benchCppPairs$ tasks), and \benchJavaTools{} Java analysis tools paired with \benchJavaProjects{} Java projects ($\benchJavaTools \times \benchJavaProjects = \benchJavaPairs$ tasks), for a total of \benchTotalTasks{} tasks.
Each task is executed independently in a fresh container. Although tasks may share the same project or analysis tool (for example KLEE+curl shares the same project with the task CSA+curl), we treat each tool-project pair as a distinct setup-and-analysis task.

\subsection{Validation of Success}
\label{subsec:manual_ground_truth}

A central challenge in evaluating agentic automation of software analysis is distinguishing genuine end-to-end analysis from superficial ``success'' signals.
For example, an agent might install an analysis tool, run its \texttt{--help} command, and then declare success without actually analyzing the target project.
To thoroughly check for meaningful analysis, we manually construct reference artifacts and validation mechanisms for each tool-project pair.

\paragraph{\textbf{Reference setups and artifacts}}
For each $(T, P)$ pair, we manually construct a containerized reference setup that installs~$T$, clones~$P$, performs the required build steps, and runs the analysis under the benchmark budget, producing a reference evidence package: full logs, key build outputs, and tool-specific analysis artifacts (e.g., static analysis reports, call graphs, generated tests, fuzzing queues, or JVM diagnostic snapshots). These artifacts define what a meaningful execution looks like for the $(T,P)$ pair.

\paragraph{\textbf{Validation process}}
When an agent completes a task, we re-execute the agent's produced setup (e.g., Dockerfile and scripts) to confirm reproducibility, and compare the agent's artifacts against the reference evidence package.
For borderline cases, we may re-run the same produced setup for longer as a diagnostic check of whether the setup can drive the intended analysis; this does not give the agent additional repair attempts or change the produced setup.
Outlier or suspicious values (e.g., unusually low coverage, empty reports, or results inconsistent with the reference) are flagged and discussed among the authors.
To support reproducibility, our replication package includes the detailed validation protocol.
Specifically, we check:
\begin{itemize}
	\item \emph{Structural evidence:} Expected files and directories exist (KLEE's \texttt{klee-out} folder, AFL++'s \texttt{queue} directory...).
	\item \emph{Project references:} Logs or reports reference project-specific paths, symbols, or build artifacts.
	\item \emph{Semantic evidence:} Tool-specific indicators of analysis progress, such as explored paths, AFL coverage growth, generated call graphs, or static analysis warnings.
\end{itemize}
This reference-backed validation helps spot premature termination by requiring artifacts that are difficult to produce without running analysis on the target repository.

\section{Agents for Automated Analysis}
\label{sec:agents}

We evaluate four agent architectures on \benchmark to study how well LLM-based agents can perform the automated software analysis task defined in \S\ref{sec:task_formalization}.
Three are baselines adapted from prior work (\S\ref{subsec:baselines}); the fourth is \analysisagent, a custom agent we designed to address the limitations those baselines reveal (\S\ref{subsec:baseline_limitations}--\ref{subsec:analysis_agent}).
Across all agents, we provide the same task specification (\S\ref{sec:task_formalization}) and the same budget (\budgetTimeoutHours{} and \$\budgetCostCap{} per task).
The key difference is how each agent plans and executes its actions and when it decides to stop. All baseline agents rely on the LLM's own judgment to decide when to stop, without explicit output validation checks.

\subsection{Baseline Agents}
\label{subsec:baselines}

\mypara{RAG-Agent.}
Retrieval-augmented generation (RAG) is a common paradigm to enhance LLM capabilities by retrieving external knowledge~\cite{lewis2020retrieval}.
We re-implement this paradigm for the automated software analysis task in \emph{RAG-Agent}, following the design of a RAG agent described by LangChain~\cite{RAGagentLangChain} but using LiteLLM to support multiple LLM backends.
We include RAG-Agent to represent the ``plan-then-execute'' paradigm that is common in LLM-based automation pipelines. Unlike the other agents, RAG-Agent does not interactively debug inside the container; instead, it synthesizes complete scripts upfront and can only revise them based on execution feedback, without fine-grained, command-level interaction.
The agent first issues web queries to collect installation and invocation guidance, e.g., in the form of analysis tool READMEs, GitHub issues, and tutorials.
Then, the agent synthesizes executable artifacts: a Dockerfile, a setup script that installs dependencies and builds the project, and a launch script that runs the analysis. To run it on an analysis task, we seed RAG-Agent with the analysis tool name, a link to the repository, the project name and URL.

\mypara{Mini-SWE-Agent.} Mini-SWE-Agent is an agent designed for software engineering tasks~\cite{Yang2024a,Ma2024}.
Instead of producing a single large script, it alternates between reasoning and tool actions (shell commands, file edits), updating its plan based on observed errors.
This style is well suited to incremental debugging and refinement.
While conceptually simple, Mini-SWE-Agent has demonstrated strong performance on software engineering tasks and is one of the leading openly available agents on SWE-bench Verified~\cite{Jimenez2024SWEBench}. For our study, we modify the prompt to specify the automated software analysis task.

\mypara{ExecutionAgent.}
ExecutionAgent~\cite{issta2025_ExecutionAgent} is an environment-setup agent designed to build a given project and execute its test suite in a container environment.
Compared to Mini-SWE-Agent, it prioritizes creating a runnable container specification and then executing the workload within that environment. We disable ExecutionAgent's test-specific retrieval mechanisms (e.g., test-suite discovery and CI workflow parsing), which are not applicable to the analysis task.

\subsection{Limitations and Failure Trends}
\label{subsec:baseline_limitations}

While the baselines differ in planning and execution style, applying them to automated software analysis reveals three common limitations.

\mypara{1) Missing procedural structure leads to stage mixing.}
All baselines occasionally attempt operations in the wrong order, such as cloning a repository into a non-existent container context or invoking analyzers before the analysis tool's dependencies are installed or before project build artifacts exist.
Because error messages are often locally scoped (e.g., missing binary, missing path, missing library), agents frequently respond with additional installation attempts instead of correcting the underlying ordering mistake.

\mypara{2) Verbose logs obscure root causes, impeding recovery.}
Long logs produced when executing commands may obscure the root cause of a failure.
RAG-Agent in particular suffers from this problem due to its monolithic scripts, but Mini-SWE-Agent and ExecutionAgent also occasionally face this challenge.
In all cases, poor error localization reduces the agent's ability to form targeted hypotheses (e.g., ``the project uses CMake but container is missing \texttt{ninja-build}''), wasting cycles on uninformed trial and error.

\mypara{3) Weak success validation causes premature termination.}
All baselines sometimes stop after observing superficial signals of success. Common examples include:
\begin{itemize}
	\item \emph{Tool presence only:} Commands like \texttt{klee -{}-version} or \texttt{infer -{}-help} succeed, so the agent declares success.
	\item \emph{Toy execution:} The analysis tool runs on a trivial input (e.g., a small file or example program), but not on the target project.
	\item \emph{Partial pipeline:} The project builds successfully, but the analysis is skipped or misconfigured, producing no tool-specific results.
\end{itemize}

\paragraph{Example 1: Stage mixing and environment thrashing (Mini-SWE-Agent, CSA + curl, Gemini-3-Flash).}
Mini-SWE-Agent correctly reports 3 warnings, yet the run is recorded as a failure.
Stage mixing causes the agent to interleave package installation with analysis attempts, wasting 17 of 25 iterations on \texttt{dpkg} lock contention from concurrent package-manager invocations.
Although the analysis does complete, a stale Docker-cleanup status message misleads the agent into believing the environment has failed, so it never recognizes its own success.
This example illustrates how \emph{stage mixing} (limitation~1) wastes cycles and how \emph{weak success validation} (limitation~3) causes the agent to discard valid results.

\paragraph{Example 2: Repetitive cycling without progress (RAG-Agent, WALA + Checkstyle, DeepSeek-V3.2).}
RAG-Agent uses all 120 iterations trying to build WALA from source (Maven, Gradle, different JDKs, hallucinated version tags) without ever diagnosing the root cause (missing C++ toolchain for WALA's \texttt{cast} module).
This example illustrates \emph{poor error localization} (limitation~2).

\subsection{\analysisagent}
\label{subsec:analysis_agent}

\begin{figure}
	\centering
	\includegraphics[width=\columnwidth]{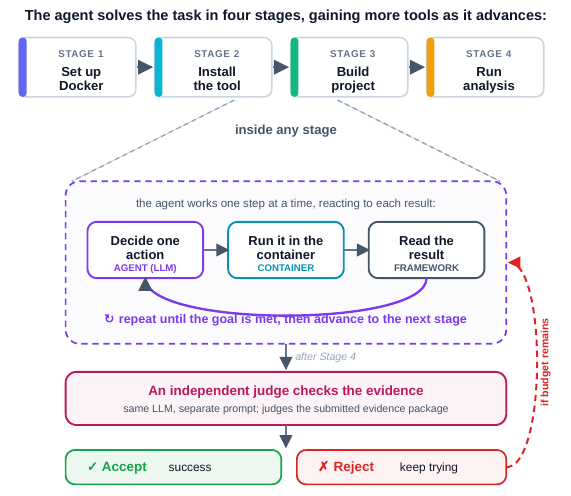}
	\vspace{-2em}
	\caption{Overview of \analysisagent. The agent advances through four stages; within any stage it loops over single actions until the goal is met, and an independent judge validates the final evidence (\S\ref{subsec:analysis_agent}).}
	\label{fig:architecture}
\end{figure}

Motivated by these limitations, we introduce \analysisagent, a custom agent built on three principles (Figure~\ref{fig:architecture}): staged execution, single-action cycles with log condensation, and evidence-based validation.
\analysisagent departs from prior agents in two ways: it enforces a fixed analysis workflow through stage-specific action constraints, rather than leaving ordering to the model as in ReAct~\cite{Yao2023ReAct} or Reflexion~\cite{Shinn2023Reflexion}; and it explicitly validates whether the produced artifacts constitute a project-level analysis result, rather than relying on successful command execution or the agent's own completion claim as in ExecutionAgent~\cite{issta2025_ExecutionAgent}.

\subsubsection{Principle 1: Explicitly Staged Execution}
The agent follows a four-stage linear progression with per-stage instructions:
\begin{enumerate}
\item \emph{Docker setup}. Writes a Dockerfile and auxiliary files. No container exists yet, so terminal commands are disabled.
  The framework enforces a tool whitelist, allowing only \texttt{write\_to\_file} and \texttt{read\_file}; any other action is rejected. The agent designs a compatible base image and installs build dependencies.

	\item \emph{Analysis tool setup}. The framework automatically advances to this stage once the container is running and responds to a shell command. In this stage, we instruct the agent to install the analysis tool and conduct a successful smoke test of the analysis tool, e.g., by invoking it on a small toy example or showing \path{--help}.

	\item \emph{Project setup}. We give the agent instructions to build the project and generate tool-specific artifacts (e.g., compilation database for static analyzers or bitcode for KLEE).
	We instruct the agent to check the availability of such artifacts before advancing to the next stage.

	\item \emph{Analysis run}. In this stage, the agent is instructed to apply the installed analysis tool to the project (i.e., not only toy examples) and provide the outputs in a results directory. When the agent declares the task complete, an LLM-as-judge validates the claim; if the judge rejects, execution continues if cycles remain (Principle~3).

\end{enumerate}

The staged design also separates reusable tool setup from project-specific work: in principle, the Dockerfile and analysis-tool installation from Stages~1--2 can be reused for additional projects that use the same tool configuration.
In our experiments, however, each task starts from scratch so that we measure end-to-end setup and analysis capability for each tool-project pair.

Only the Stage~1$\to$2 transition is framework-driven: the framework advances once the container responds to a shell command, since the agent cannot yet observe the container.
The Stage~2$\to$3 and Stage~3$\to$4 transitions are agent-driven: the agent emits a \texttt{STAGE\_COMPLETED} sentinel once smoke-test and artifact-existence checks pass, respectively.

In each stage, the prompt carries over a summary of what happened in previous stages (e.g., which versions of the analysis tool and its dependencies were installed, compilation results, preview of latest actions).
The carry-over information contains recent commands from completed stages and a synthesized summary obtained by a separate LLM query.
While staged pipelines are common in CI/CD, our decomposition is tailored to automated software analysis.

\subsubsection{Principle 2: Single-Action Cycles with Deterministic Log Condensation}
\analysisagent separates reasoning from command formatting because weaker LLMs struggle to produce correctly structured tool calls inline.
It therefore triggers actions in two steps:
First, the agent receives stage context, recent observations, and structured feedback from prior failures, and outputs free-form reasoning with proposed next steps. Second, a separate LLM call extracts the first concrete action from this response, formats it as an executable command (following a per-tool JSON schema that we define for each agentic tool), and the system executes it, returning the observation (stdout/stderr, exit code, environment metadata) for the next cycle.
Invalid actions (e.g., stopping the container, host-to-container copies) are rejected by the framework with an error message returned to the agent, and the cycle is consumed without effect.
Because build systems and analysis tools can produce thousands of lines of output, \analysisagent applies deterministic pattern matching to condense logs into diagnostic signals, unlike prior work that relies on LLM-based summarization~\cite{issta2025_ExecutionAgent}: the first \logHeadChars{} characters (command + initial progress), lines matching any of \logFailurePatterns{} manually curated failure patterns (e.g., compiler/linker errors, missing files, package errors), and the last \logTailChars{} characters (exit status + summary).
The action-observation loop follows the ReAct paradigm~\cite{Yao2023ReAct}; the key differences are the separation of planning from extraction and the deterministic log condensation, both tailored to the long, noisy outputs of build systems and analysis tools.

\subsubsection{Principle 3: Evidence-Based Validation}
When the agent declares the task complete, the system constructs an \textit{evidence package} comprising three components: (1)~stage summaries, including successful and failed commands and environment choices, (2)~recent analysis-stage observations, and (3)~output file locations.
This package is then submitted to a separate LLM-as-judge validation call.
The judge uses the same LLM backend as the agent, but receives only the structured evidence package rather than the full conversation or the agent's reasoning.
The judge is thus isolated through a separate prompt and restricted context rather than a different model backend; we evaluate backend separation separately through the \emph{cross-backend judge} ablation in \S\ref{subsubsec:rq_ablation}.
The judge verifies that (i)~the invocation targets the specified project and (ii)~output artifacts consistent with analysis tool completion exist (e.g., KLEE test cases, AFL++ crash queues).
It then compares the actual output against a reference example that an LLM synthesizes from the analysis tool's official documentation once per tool (reused across all tasks for that tool), which helps exclude trivial or error-produced output.
If validation fails, the system notifies the agent of the rejection reason and triggers continued execution (if cycles remain).
This LLM-as-judge serves as an internal heuristic to reduce premature termination. We evaluate the judge's agreement with manual verification in \S\ref{subsubsec:rq_ablation}.
While LLM-as-judge approaches have been used for general-purpose evaluation~\cite{Zheng2024}, the structured evidence package and tool-specific reference comparison target the challenge of distinguishing genuine analysis completion from superficial success.

\section{Evaluation}
\label{sec:evaluation}
We organize the evaluation around five research questions:

\begin{itemize}[leftmargin=*,itemsep=1pt]
	\item \textbf{RQ1 (Effectiveness):} How effective are agents on \benchmark across different configurations?
	\item \textbf{RQ2 (Qualitative analysis):} What distinguishes successful from unsuccessful runs, and what failure modes occur?
	\item \textbf{RQ3 (Efficiency and costs):} What are the time, cycle, and cost profiles for different agents?
	\item \textbf{RQ4 (Ablation):} How much does each of \analysisagent's three principles contribute to verified success?
	\item \textbf{RQ5 (Extended runs):} Can agent-produced setups support meaningful analysis beyond benchmark run?
\end{itemize}

\subsection{Experimental Setup}
\label{subsec:exp_setup}

\begin{table}[t]
	\centering
	\caption{LLMs used in evaluation (API prices per 1M tokens, March 2026).}
	\label{tab:llm_backends_pricing}
	\resizebox{\columnwidth}{!}{%
	\begin{tabular}{lcccc}
		\toprule
		\textbf{Model} & \textbf{Input price} & \textbf{Output price} & \textbf{in/out limits}\\
		\midrule
		\texttt{gpt-5-nano} & \$0.05 & \$0.40 & 400K / 128K \\
		\texttt{gpt-5-mini} & \$0.25 & \$2.00 & 400K / 128K \\
		\texttt{deepseek-v3.2} & \$0.56 & \$1.50 & 128K / 64K \\
		\texttt{gemini-3-flash} & \$0.50 & \$3.00 & 1M / 64K \\
		\bottomrule
	\end{tabular}%
	}
\end{table}

\paragraph{\textbf{Configuration.}}
We evaluate all agents on the full \benchmark benchmark of \benchTotalTasks{} tasks (\S\ref{subsec:benchmark_dataset}) using four LLM backends of varying capability and cost (Table~\ref{tab:llm_backends_pricing}).
All agents share the same per-task stopping limits: a maximum of \budgetMaxCycles{} agentic cycles, a \$\budgetCostCap{} API cost cap, and a \budgetTimeoutHours{} wall-clock timeout; reaching any limit interrupts the task and marks it as failed.
For agents supporting retries (\analysisagent, ExecutionAgent), each retry counts as additional cycles within the same per-task budget.
We repeat each configuration $n=3$ times to account for non-determinism.

\begin{table*}[t]
    \centering
	\caption{Self-validated and manually verified success rates by agent architecture and LLM backend. Self-validated rates are mean $\pm$ standard deviation over $n{=}3$ runs; manually verified rates are from a single run.}
    \label{tab:eval_results}
    \resizebox{0.9\textwidth}{!}{%
    \begin{tabular}{l|cc|cc|cc|cc|c}
        \toprule
        & \multicolumn{2}{c|}{\textbf{GPT-5-nano}} & \multicolumn{2}{c|}{\textbf{GPT-5-mini}} & \multicolumn{2}{c|}{\textbf{DeepSeek-V3.2}} & \multicolumn{2}{c|}{\textbf{Gemini-3-Flash}} & \\
		\textbf{Agent} & Self-val. & Verified & Self-val. & Verified & Self-val. & Verified & Self-val. & Verified & \textbf{Avg.\ Verified} \\
        \midrule
        RAG-Agent            & $\phantom{0}73 \pm \phantom{0}7$\% & \phantom{0}9\% & $\phantom{0}98 \pm \phantom{0}1$\% & \phantom{0}6\% & $\phantom{0}49 \pm 16$\% & \phantom{0}3\% & $\phantom{0}93 \pm \phantom{0}5$\% & 23\% & 10\% \\
        Mini-SWE-Agent       & $100 \pm \phantom{0}0$\% & \phantom{0} 9\%   & $100 \pm \phantom{0}0$\% & 20\% & $100 \pm \phantom{0}0$\% & 57\% & $\phantom{0}87 \pm 16$\% & 63\% & 37\% \\
        ExecutionAgent       & $\phantom{0}30 \pm 37$\% & 40\% & $\phantom{0}63 \pm 31$\% & 54\% & $\phantom{0}50 \pm \phantom{0}3$\% & 57\% & $\phantom{0}86 \pm 19$\% & 77\% & 57\% \\
        \analysisagent       & $\phantom{0}77 \pm 21$\% & 54\% & $\phantom{0}97 \pm \phantom{0}4$\% & 75\% & $94 \pm \phantom{0}3$\% & 91\% & $97 \pm \phantom{0}2$\% & 94\% & 79\% \\
        \bottomrule
    \end{tabular}%
    }
\end{table*}

\paragraph{\textbf{Success metrics.}}
We assess task outcomes with two complementary metrics.
First, a task is \emph{self-validated} if the agent terminates and claims task completion. For \analysisagent, this includes passing its internal LLM-as-judge validation (\S\ref{subsec:analysis_agent}); for baseline agents, it means the agent exited without error and reported success. Self-validated success may include false positives.
Second, a task is \emph{manually verified} if it also passes the manual validation process described in \S\ref{subsec:manual_ground_truth}. We manually verify only one run of the three.

\paragraph{\textbf{Statistical testing.}}
We test whether the observed advantage of \analysisagent{} over each baseline is larger than would be expected from random variation. For self-validated success, we compare agents task by task: when \analysisagent{} and a baseline have a different outcome on a task (e.g., fail vs.\ success), McNemar's one-sided exact test~\cite{mcnemar1947note} tests whether the disagreement favors \analysisagent{} more often than the baseline. For manually verified success, we compare the numbers of verified successes and failures in the manually inspected runs using Fisher's one-sided exact test. In both cases, our hypothesis is that \analysisagent{} solves more tasks than each baseline (with significance). Since we make multiple pairwise comparisons, we control the family-wise error rate with Holm-Bonferroni correction~\cite{holm1979simple} at $\alpha{=}0.05$: 12 per-backend comparisons for self-validated success (3 baselines $\times$ 4 LLMs corresponding to colored rows in Figure~\ref{fig:forest_plot}) and 3 pooled comparisons for manually verified success (one per baseline; black rows in Figure~\ref{fig:forest_plot}).

\subsection{RQ1: Effectiveness Across Agents and LLMs}
\label{subsubsec:rq1}

The gap between self-validated and manually verified rates is large across all agents (Table~\ref{tab:eval_results}), underscoring why LLM-internal completion claims cannot replace artifact-based verification.

\paragraph{\textbf{Verified success.}}
Manually verified success rates span a wide range across agent architectures: RAG-Agent averages only \resRAGAvgVerified\% across LLM backends (range 3--23\%), showing that naive retrieval-based prompting is insufficient for multi-step analysis workflows.
Mini-SWE-Agent, a general-purpose coding agent, reaches an average of \resSWEAvgVerified\% but exhibits high LLM sensitivity (9--63\%), suggesting that without task-specific scaffolding, effectiveness depends heavily on the model's capabilities.
ExecutionAgent's tool-aware adaptations raise the average to \resExecAvgVerified\%, with less variation across LLMs (40--77\%).
\analysisagent achieves the highest manually verified success, averaging \resAAAvgVerified\% across LLM backends (range 54--94\%).
The best configuration, \analysisagent with Gemini-3-Flash, achieves \bestVerifiedRate\% verified success (\bestVerifiedCount{} / \benchTotalTasks{} tasks); DeepSeek-V3.2 reaches 91\%, showing that purpose-built scaffolding reliably handles multi-step analysis workflows across different backends.

\begin{figure*}
	\centering
	\includegraphics[width=1\textwidth]{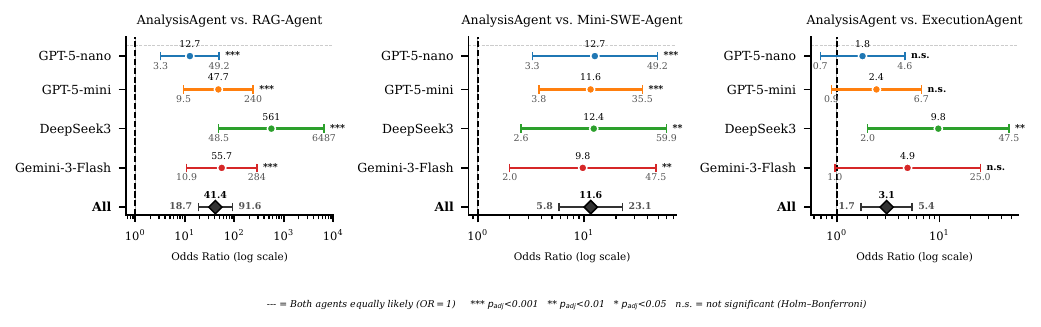}
	\vspace{-1.75em}
	\caption{How many times more likely \analysisagent is to succeed compared to baselines, per LLM backend (circles) and pooled across all LLMs (diamonds). Values above~1 mean \analysisagent succeeds more often; 1 means equal odds.}
	\label{fig:forest_plot}
\end{figure*}

\emph{\textbf{Statistical analysis.}}
In short, \analysisagent's advantage over all three baselines is large, consistent across all LLM backends, and statistically significant.
Figure~\ref{fig:forest_plot} visualizes the per-LLM success odds ratio of \analysisagent vs.\ other agents. We organize the statistical evidence around two questions:
\begin{enumerate}[label=(\arabic*)]

\item \emph{Does \analysisagent outperform the baselines, and by how much?}
Fisher's exact test with Holm-Bonferroni correction yields $p_{\mathrm{adj}}{<}0.001$ for all three pairwise comparisons, pooled across LLM backends. The effect sizes are large: Cohen's $h{=}1.55$ vs.\ RAG-Agent, $h{=}0.92$ vs.\ Mini-SWE-Agent, and $h{=}0.45$ vs.\ ExecutionAgent.
The odds ratios are $\mathrm{OR}{=}34.5$ [17.3, 68.5] vs.\ RAG-Agent, $\mathrm{OR}{=}8.1$ [4.0, 16.2] vs.\ Mini-SWE-Agent, and $\mathrm{OR}{=}2.7$ [1.6, 4.6] vs.\ ExecutionAgent; all 95\% confidence intervals exclude~1.

\item \emph{Does the advantage hold across all LLM backends?}
The Cochran-Mantel-Haenszel test controls for LLM choice and confirms significance for all three baselines ($\chi^2_{\mathrm{MH}}{=}139.8$, $p{<}10^{-10}$; $43.3$, $p{=}4.7{\times}10^{-11}$; $15.4$, $p{=}8.8{\times}10^{-5}$), with common odds ratios 41.4, 11.2, 3.1.

\end{enumerate}

\emph{\textbf{Self-validated success.}}
Self-validated success paints a misleadingly optimistic picture. McNemar's exact test shows that \analysisagent significantly outperforms ExecutionAgent on all backends ($p_{\mathrm{adj}}{<}0.001$), but not RAG-Agent or Mini-SWE-Agent, because those agents frequently claim completion even when the submitted artifact is incorrect (e.g., Mini-SWE-Agent on DeepSeek-V3.2 reports 100\% self-validated success but only 57\% verified success). The gap between self-validated and manually verified success is smallest for \analysisagent, especially with Gemini-3-Flash (\bestSelfValidatedRate\% vs.\ \bestVerifiedRate\%), and largest for RAG-Agent and Mini-SWE-Agent. This pattern indicates that false positives are driven by both model limitations and agent architecture: weaker models terminate prematurely more often, but model capability alone does not compensate for agents that lack explicit artifact validation, stage discipline, and error-localized recovery.

\paragraph{\textbf{Tool and ecosystem patterns.}}
\begin{figure}
	\centering
	\includegraphics[width=0.85\columnwidth]{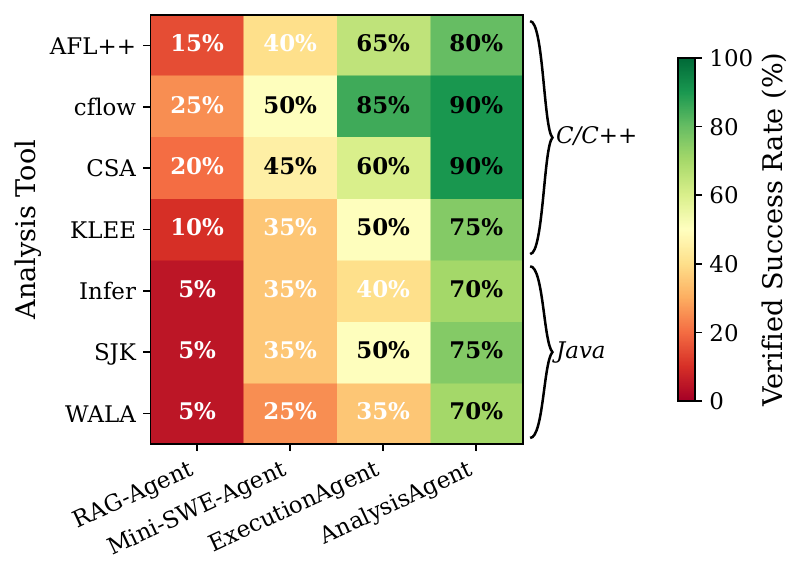}
	\caption{Average verified success rate by analysis tool and agent, aggregated over all LLM backends.}
	\label{fig:success_by_tool}
\end{figure}

Figure~\ref{fig:success_by_tool} stratifies manually verified success rates by analysis tool and agent.
Across all agents, cflow achieves the highest average success rate (\toolSuccessCflow\%), followed by CSA (\toolSuccessCSA\%) and AFL++ (\toolSuccessAFL\%), while WALA has the lowest success rate (\toolSuccessWALA\%).
WALA and Infer are the largest failure contributors (\toolFailShareWALA\% and \toolFailShareInfer\% of total failures, respectively), followed by SJK (\toolFailShareSJK\%) and KLEE (\toolFailShareKLEE\%).
At the ecosystem level, Java tasks account for \failJavaShareOfFailures\% of all failures versus \failCppShareOfFailures\% for C/C++, and this gap persists across all agents, reflecting the complexity of Java toolchains (classpaths, bytecode generation, JVM attachment) and heavyweight whole-program analyses.

\subsection{RQ2: Qualitative Analysis}
\label{subsubsec:rq_qualitative}

We describe success patterns in \analysisagent runs, then classify all 182 failed trajectories across all four agents into root-cause categories derived by iterative open coding of cycle-level traces. Baseline failure trends are also discussed in \S\ref{subsec:baseline_limitations}.

\subsubsection{Success Patterns}
Successful \analysisagent runs share three characteristics:
(i)~early convergence on a working container (typically within \cycConvergenceCycles{} cycles),
(ii)~incremental stage-by-stage progress with targeted error recovery, and
(iii)~evidence production and validation at each stage.

\paragraph{Example: KLEE on masscan (DeepSeek-V3.2)}
The agent applies KLEE to masscan, a C port scanner, in 29 cycles (\$0.39, 12\,min).
After spending 14~cycles on dependency resolution (TCMalloc, SQLite3, klee-uclibc), it identifies \texttt{massip\_parse\_ipv4} as an entry point, compiles to LLVM bitcode, and runs KLEE, producing 28~completed and 172~partial test paths confirmed by the validator.

\begin{table}[t]
\centering
\caption{Analysis output statistics for self-validated successful \analysisagent{} runs across all three repetitions (377/420).}
\label{tab:tool_output_stats}
\resizebox{\columnwidth}{!}{%
\begin{tabular}{@{} l l r@{--}l r r @{}}
\toprule
Analysis tool & Metric & \multicolumn{2}{c}{IQR} & Median & Time (s) \\
\midrule
\multirow{4}{*}{AFL++} & Code Coverage (\%) & 0.39 & 1.2 & 0.44 & \multirow{4}{*}{210} \\
 & Tested Inputs & 19K & 1.8M & 230K &  \\
 & Exec Speed Per S & 108 & 1.6K & 569 &  \\
 & Corpus Size Seeds & 0 & 63 & 6 &  \\
\addlinespace[2pt]
\multirow{3}{*}{KLEE} & Test Cases Generated & 1 & 32 & 2 & \multirow{3}{*}{17} \\
 & Completed Paths & 0 & 6 & 1 &  \\
 & Instructions Executed & 4.3K & 95K & 13K &  \\
\addlinespace[2pt]
CSA & Bugs Reported & 0 & 24 & 0 & 144 \\
\addlinespace[2pt]
\multirow{2}{*}{cflow*} & Output Lines & 25 & 118 & 60 & \multirow{2}{*}{103} \\
 & Functions Listed & 11 & 69 & 41 &  \\
\addlinespace[2pt]
Infer & Issues Reported & 3 & 24 & 3 & 102 \\
 & \multicolumn{5}{@{}l}{\footnotesize \hspace{.5em} Top: thread safety (283), resource leak (16), null deref (13)} \\
\addlinespace[2pt]
\multirow{3}{*}{WALA} & Call Graph Nodes & 12K & 157K & 39K & \multirow{3}{*}{204} \\
 & Cha Classes & 7.2K & 29K & 9.5K &  \\
 & Call Graph Edges & 148K & 5.2M & 4.3M &  \\
\addlinespace[2pt]
\multirow{2}{*}{SJK} & Process Cpu (\%) & 0 & 1K & 101 & \multirow{2}{*}{66} \\
 & Thread Count & 10 & 59 & 15 &  \\
\bottomrule
\end{tabular}%
}
\par\vspace{2pt}{\footnotesize *cflow row reflects the truncated output the agent sees (full output is larger).}
\end{table}

\subsubsection{Analysis Outputs}

Beyond binary success, we examine the artifacts produced by successful \analysisagent{} runs. Table~\ref{tab:tool_output_stats} summarizes tool-specific output metrics using medians and inter-quartile ranges (IQRs). Successful runs produce concrete evidence across all seven tools; for example, Infer reports thread-safety, resource-leak, and null-dereference issues, while WALA constructs call graphs with a median of 39K~nodes and 4.3M~edges. Dynamic tools remain intentionally shallow under the benchmark budget: AFL++ reaches a median coverage of 0.44\% and KLEE generates a median of two test cases because the agent is instructed to spend only 30--180\,s on final analysis invocations.

\subsubsection{Failure modes}
Of the 560 manually reviewed agent--backend--task instances, we coded 182 failed trajectories with complete cycle-level traces, excluding those with infrastructure crashes and missing logs. The categories are non-exclusive:
\begin{itemize}
    \item \emph{Docker/build failure} (67): The agent is stuck in a Docker-build loop, fails to install the analysis tool, or cannot compile the target project, never reaching the analysis stage.
	\item \emph{Analysis tool misuse} (73): The tool is installed, but the agent invokes it incorrectly (wrong API classes, stale imports, wrong flags or input formats) or cannot produce required prerequisites, such as LLVM bitcode or a fuzzing harness.
    \item \emph{Malformed LLM output} (42): The LLM produces output that the agent framework cannot parse as a valid tool call, causing cycles to be consumed without any action.
    \item \emph{Budget/time exhausted} (34): The iteration budget or wall-clock limit is exhausted (e.g., while stuck in Docker build).
    \item \emph{Incorrect analysis result} (21): The agent completes execution and submits a result, but the output is rejected by the validator or found to be incorrect during manual validation.
\end{itemize}

Each agent exhibits a distinct failure profile.
RAG-Agent and ExecutionAgent primarily fail during environment construction:
RAG-Agent is dominated by Docker/build failures (60\%), reflecting its tendency to exhaust all 120 Dockerfile-generation iterations without learning from repeated build errors;
ExecutionAgent is dominated by malformed LLM output (50\%), followed by Docker/build failures (20\%). The malformed-output failures are concentrated on DeepSeek-V3.2, which frequently returns empty responses that the framework cannot parse as valid tool calls, silently consuming cycles without progressing past environment setup.
Mini-SWE-Agent fails at output quality and analysis tool usage: Incorrect analysis results (48\%) and analysis tool misuse (44\%) dominate, as the agent often completes the workflow but produces inadequate analysis results or misuses the analysis tool.
Finally, \analysisagent has largely solved environment setup (only 7\% of its failures) and instead fails during analysis tool invocation: analysis tool misuse (50\%) is the leading cause, driven by incorrect WALA API usage, failed Infer build integration, and KLEE bitcode extraction failures.
Two recurring signatures stand out: 33\% of failing runs involve compile-fix loops (3+ consecutive compilation errors without resolution) and 23\% include repeated validation rejections (5+ rejections); these thresholds describe observed behavior, not tuned parameters.

For example, WALA on Closure (GPT-5-mini) spends 42~cycles on Docker/build failures before entering a prolonged analysis tool misuse loop with five cascading WALA API errors; the trajectory consumes ${\sim}$102~cycles and \$1.37, exemplifying the compile-fix loop pattern observed in 33\% of failing runs.

\subsection{RQ3: Efficiency and Costs}
\label{subsubsec:rq_efficiency}

\begin{figure}
	\centering
	\includegraphics[width=0.95\columnwidth]{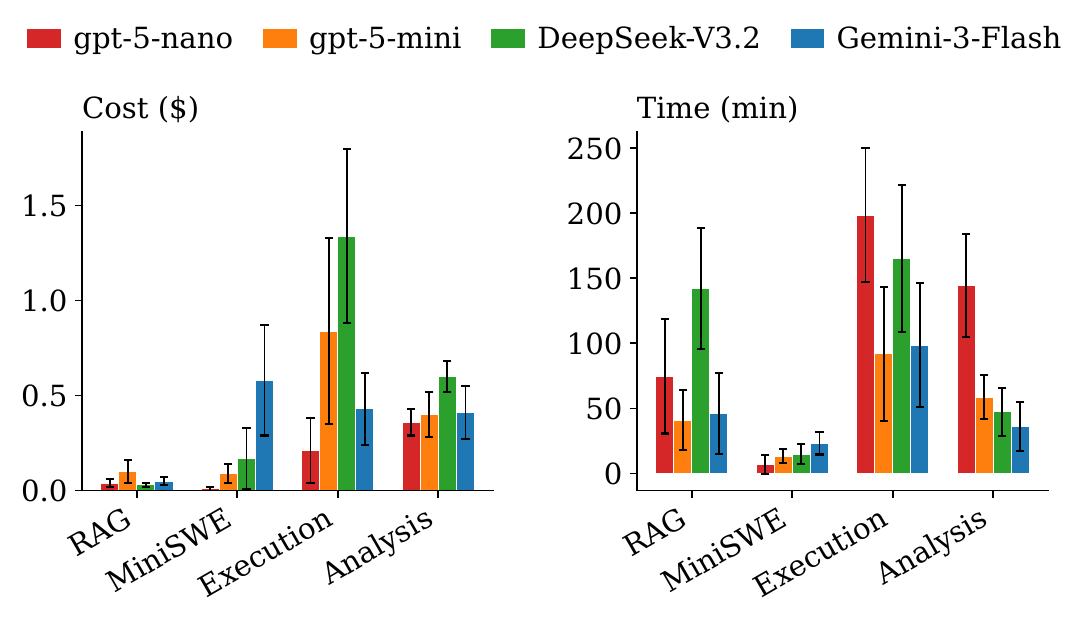}
	\vspace{-1em}
	\caption{Resource consumption by agent and LLM backend.}
	\label{fig:efficiency_comparison}
\end{figure}

\paragraph{\textbf{High variance across tasks}}
The figure shows high variance across tasks, with standard deviations often large relative to the means. This indicates that cost is driven not only by the agent--model pair, but also by task-specific factors such as dependency resolution, build-system complexity, retry behavior, and early failure modes.

\paragraph{\textbf{Capability-efficiency interaction}}
Model capability and efficiency are tightly coupled.
The weakest backend (GPT-5-nano) consumes the most calls and wall-clock time (mean \cycNanoMeanCalls{} calls and \timemNanoMean{} minutes per task), despite having the lowest per-token pricing.
Stronger backends substantially reduce interaction length and execution time: DeepSeek-V3.2 achieves a mean of \cycDeepSeekMeanCalls{} calls and \timemDeepSeekMean{} minutes per task, and Gemini-3-Flash achieves a mean of \timemGeminiMeanMin{} minutes per task.
Hence, token price alone is not a reliable proxy for per-task cost, as weaker models require more cycles and more tool executions.

\paragraph{\textbf{Failure cost and model sensitivity}}
Failed runs are consistently more expensive than successful runs, with \effFailCycleMult$\times$ more cycles, \effFailDurationMult$\times$ longer duration, \effFailCostMult$\times$ higher cost, and \effFailRetryMult$\times$ more retry attempts.
The magnitude of this inflation varies by model: Gemini-3-Flash shows the highest sensitivity (\effGeminiFailSensitivity$\times$ more cycles on failure vs.\ success) despite the lowest absolute count on success ($\sim$\cycGeminiSuccessCycles{} cycles), while DeepSeek-V3.2 degrades more gracefully (\effDeepSeekFailSensitivity$\times$).
Failures are not just unproductive: they are disproportionately expensive, consuming far more time and money than successful runs.
Choosing a weaker or cheaper model does not save cost if it also fails more often.
Similarly, reducing the timeout could curb the overhead of failed runs, but at the risk of stopping runs that would have eventually succeeded and lowering overall success rates.

\paragraph{\textbf{Language ecosystem differences}}
Java tasks are systematically more resource-intensive than C/C++ (e.g., mean \ecoJavaMeanMin{} vs.\ \ecoCppMeanMin{} minutes and \ecoJavaMeanCalls{} vs.\ \ecoCppMeanCalls{} calls for \analysisagent with GPT-5-nano; the same trend holds across all agent--model combinations), consistent with the higher failure rates reported in \S\ref{subsubsec:rq1}.
We attribute this gap to the Java analysis toolchain rather than to the language itself.
First, setup dependencies are heavier: Infer requires Buck, opam, OCaml, JDK, and Clang/LLVM, whereas C/C++ tools such as cflow require only Autotools.
Second, all three Java tools require a successful full project build before analysis can begin: Infer and WALA operate on bytecode and SJK must attach to a running JVM, while C/C++ tools such as cflow and CSA can analyze source directly or integrate into the compilation step.
Third, runtime orchestration is more complex: SJK requires launching the target as a background process and attaching the profiler, a coordination pattern absent from all C/C++ tools.

\subsection{RQ4: Ablation of Design Principles}
\label{subsubsec:rq_ablation}

To isolate the contribution of each design principle from the choice of LLM backend, we ablate each of \analysisagent's three principles in turn while holding the backend, task specification, and budget fixed.
All variants use DeepSeek-V3.2 as the agent backend and run once on the full \benchmark{} benchmark (\benchTotalTasks{} tasks).
The full-agent row is a reference run for the ablation study, and hence, not identical to the median run reported in Table~\ref{tab:eval_results}.

We evaluate four variants.
\emph{(1)~No staged execution} removes Principle~1: the agent runs in a single loop without stage splitting, per-stage instructions, or per-stage action constraints.
\emph{(2)~Raw tool output} ablates Principle~2: the agent receives raw build and analysis logs instead of deterministic diagnostic condensation.
In practice, this adds hundreds of extra log lines to the prompt in each cycle.
\emph{(3)~No LLM judge} removes Principle~3: the agent stops when it claims completion, without an additional validation step.
\emph{(4)~Cross-backend judge} changes the judge backend only from DeepSeek-V3.2 to Gemini-3-Flash, allowing us to separate backend diversity from the prompt and context isolation used by the default judge.

\begin{table}[t]
	\centering
	\caption{Ablation of \analysisagent's with DeepSeek-V3.2 (single run).}
	\label{tab:ablation}
	\setlength{\tabcolsep}{4pt}
	\begin{tabular}{lccc}
		\toprule
		\textbf{Variant} & \textbf{Self-val.} & \textbf{Verified} & \textbf{\$/task} \\
		\midrule
		Full \analysisagent (reference)          & 97\% & 89\% & \$0.34 \\
		Cross-backend judge (Gemini)             & 89\% & 80\% & \$0.36 \\
		No LLM judge                             & 77\% & 69\% & \$0.31 \\
		No staged execution                      & 74\% & 60\% & \$0.45 \\
		Raw tool output                          & 66\% & 51\% & \$0.37 \\
		\bottomrule
	\end{tabular}
\end{table}

Table~\ref{tab:ablation} confirms that each principle contributes: every ablation reduces verified success from the 89\% reference.
The largest drop comes from removing log condensation (raw-output variant: 51\%), showing that long build logs are not merely a context-window problem: even when raw output fits in the model's context window, important error messages are buried among hundreds of unrelated lines and recovery suffers.

Staged execution is the second most important component: removing it drops success to 60\%, raises average cost to \$0.45, and increases cycle count from 52 to 61, consistent with agents that pursue locally plausible repairs without restoring the intended workflow order (\S\ref{subsec:baseline_limitations}).
Removing the judge reduces success to 69\%, confirming its value as an internal stopping check, though it is not the final evaluation metric: all Verified values come from manual validation (\S\ref{subsec:manual_ground_truth}).
The cross-backend judge (80\%) underperforms the same-backend default (89\%), suggesting the gain comes from evidence-based validation with separate context rather than backend diversity.
Finally, lower-cost variants are not more efficient: removing the judge saves little but loses seven verified successes, while removing stages raises both cost and failure rate.

\subsection{RQ5: Extended-Budget Runs}
\label{subsec:extended_runs}

To assess whether the generated setups are useful beyond short-timed validation, we re-run selected successful \analysisagent{} outputs for AFL++, KLEE, and SJK under an extended \extendedRunBudget{} budget. We reuse the setup produced by \analysisagent{} (Gemini run) and set the analysis command timeout to 3 hours. Static tools are excluded because their original invocations already run to completion.

The extended runs perform substantial analyses. KLEE generates \num{386193} test cases across five projects and executes more than $4.7{\times}10^{8}$ LLVM instructions. AFL++ executes approximately $8.3{\times}10^{7}$ inputs across five projects and grows the aggregate queue to 3{,}315 entries. SJK collects 9{,}131 thread snapshots across four projects (11--21 threads each), with per-thread CPU and allocation rates peaking at 92.2\% CPU and 602\,MB/s. 
\begin{figure}[t]
\begin{lstlisting}[style=terminal,basicstyle=\ttfamily\scriptsize]
$ cmd  input value   cmd output         expected correct
$ rax2 '1<<64'    -> 0x1                (expected 0)
$ rax2 '1<<200'   -> 0x100              (expected 0)
$ rax2 '255<<255' -> 0x8000000000000000 (expected 0)
$ r2 -c 's 1<<64' -> seeks to 0x1       (expected 0)
\end{lstlisting}
\vspace{-0.5em}
{\scriptsize\textbf{(a) radare2 (KLEE): wrong result for over-shift expressions}}

\medskip
\begin{lstlisting}[style=terminal,basicstyle=\ttfamily\scriptsize,escapeinside={(*@}{@*)}]
$ cat crash.conf            # AFL-generated input
rate 1
ports = i:53
router- = 80
range = 7.0.0.7-020.10.0.12

$ masscan -c crash.conf --echo
[-] CONF: bad MAC address: router- = 80
(*@\textcolor{red}{Bus error (core dumped)   \# exit 135 (SIGBUS)}@*)
\end{lstlisting}
\vspace{-0.5em}
{\scriptsize\textbf{(b) masscan (AFL++): crash on malformed scan configuration}}

\caption{Two confirmed defects surfaced by KLEE and AFL++ in longer runs.}
\label{fig:extended_bugs}
\end{figure}

In particular, both AFL++ and KLEE report crashes and errors with one confirmed bug each (Figure~\ref{fig:extended_bugs}). The KLEE finding is in radare2's numeric-expression evaluator, \texttt{r\_num\_calc}, which is used by \texttt{rax2} and by the \texttt{r2} command line for numeric arguments, such as addresses, sizes, and seek offsets. KLEE's symbolic command-line argument drives the evaluator to shift a 64-bit value by at least the operand width. Such shifts are undefined in C and should be rejected or normalized by the evaluator. Instead, the unmasked shift count produces incorrect values, e.g., \texttt{rax2 '1<{}<64'} returns \texttt{0x1} rather than \texttt{0}. Because the same evaluator is used inside \texttt{r2}, the defect also affects commands such as \texttt{r2 -c 's 1<{}<64'}, which seeks to \texttt{0x1} rather than \texttt{0}.
We reported this defect to the radare2 maintainers.\footnote{\url{https://github.com/radareorg/radare2/issues/26359}}

The AFL++ finding is in masscan's scan-range and configuration parser. The generated configuration is malformed and should be rejected. Instead, parsing the malformed config corrupts an internal range-list pointer with bytes derived from the input. The program later dereferences the corrupted pointer while growing the list in \texttt{rangelist\_add\_range}, causing a SIGBUS crash. We reproduced the crash on a clean build, confirming that it is a memory-safety defect.
We reported this defect to the masscan maintainers.\footnote{\url{https://github.com/robertdavidgraham/masscan/issues/853}} Overall, the extended runs show that the agent-produced setups support real downstream analysis and can surface new unknown defects.

\section{Threats to Validity}
\label{sec:threats}

\benchmark{} covers \benchNumTools{} analysis tools and \benchNumProjects{} projects, and therefore does not span the full diversity of software-analysis scenarios. Our scope is deliberate: We focus on command-line targets with well-defined entry points to separate setup difficulty from harness construction, open-source tools to support reproducibility, and C/C++ and Java because both ecosystems provide mature analysis infrastructure. Generalizing to libraries, commercial analyzers, and additional ecosystems, such as Python or Rust, remains future work. The \benchTotalTasks{} tasks are also not statistically independent, since multiple tasks share projects and tools. With $n{=}3$ repetitions, variance estimates are necessarily rough, especially for weaker models with high non-determinism. We therefore report standard deviations and avoid drawing fine-grained conclusions from them.

Manual validation was performed by a single primary validator, who re-executed agent setups and compared the resulting artifacts against the task-specific reference criteria. Outlier cases were discussed among the authors, but we did not measure formal inter-rater agreement. However, the validation procedure relies primarily on reproducible artifact checks, such as file existence, project-specific references, tool output, and analysis evidence. We include the verification protocol and logs to support independent inspection.

Our success criterion uses tool-specific validity checks rather than numeric thresholds such as minimum AFL++ coverage; low dynamic-analysis medians in Table~\ref{tab:tool_output_stats} therefore reflect short benchmark runtimes, not poor setup quality.

\section{Related Work}
\label{sec:related_work}

\paragraph{Environment construction and repository-level agents}
Reliable software construction is a long-standing challenge~\cite{lou2020understanding}. Prior work has studied dependency repair for reproducible Python builds~\cite{mukherjee2021fixing}, learning-based localization and repair of Java build errors~\cite{Tarlow2020}, reproducibility problems in bug datasets~\cite{Zhu2023} and computational notebooks~\cite{wang2020assessing}, and dependency inference for executable code snippets, as in DockerizeMe~\cite{horton2019dockerizeme}. More recently, ExecutionAgent introduced LLM-based environment setup for repository-level execution~\cite{issta2025_ExecutionAgent}, with follow-up work exploring related agents and SWE-bench-like benchmarks for repository tasks~\cite{Milliken2025,Eliseeva2025,Yang2025,hu2025llm,Jimenez2024SWEBench,vergopoulos2025automated}. These systems address the important problem of making a project build or run. Our setting is broader: the agent must also install an external analysis tool, satisfy tool-specific prerequisites such as bitcode, compilation databases, classpaths, or running JVMs, invoke the tool correctly, and produce verifiable analysis artifacts. This additional analysis layer is the main distinction from prior environment-setup agents.

\paragraph{LLM-assisted software analysis and software engineering agents}
Several systems use LLMs to support specific software-analysis tasks. ChatAFL~\cite{Meng2024ChatAFL} and FuzzGPT~\cite{Deng2024FuzzGPT} use the LLM as a component within a fixed, tool-specific analysis workflow. In contrast, \analysisagent{} applies a heterogeneous set of analysis tools, including fuzzers, symbolic executors, static analyzers, profilers, and structural analyzers, end-to-end across C/C++ and Java projects. More broadly, LLM-based agents have been applied to issue resolution and program repair~\cite{Yang2024a,icse2025-RepairAgent,Zhang2024a,Wang2024a,Lee2024,Huang2024,Tao2024,Liu2024a,gao2025trae}, test generation~\cite{Muendler2024,Ahmed2024,icse2026_Issue2Test,Cheng2025}, root-cause analysis~\cite{Roy2024}, and notebook debugging~\cite{Grotov2024}. These tasks usually operate inside an existing software environment; our task requires the agent to construct the environment, configure an additional analysis toolchain, and validate that the resulting artifacts are meaningful rather than merely executable.

\paragraph{Agent design and benchmarks}
\analysisagent{} builds on general ideas from agentic prompting. ReAct interleaves reasoning and action~\cite{Yao2023ReAct}, while Reflexion uses verbal self-reflection to improve subsequent decisions~\cite{Shinn2023Reflexion}. Our design is complementary: it uses short feedback-driven cycles, but structures them around staged execution, deterministic log condensation, and evidence-based validation. Existing benchmarks cover issue solving~\cite{Jimenez2024SWEBench}, interactive agent tasks~\cite{liu2024agentbench}, research-repository execution~\cite{Bogin2024}, scientific-result replication~\cite{Siegel2024,Hu2025,Starace2025}, and program-analysis tool evaluation~\cite{hazimeh2020magma,metzman2021fuzzbench}. \benchmark{} differs by combining repository setup with external tool installation, tool-specific artifact preparation, analysis invocation, and validation of the produced evidence. This combination is not captured by existing benchmarks.

\section{Conclusion}
\label{sec:conclusion}

We introduce the task of automated software analysis and present \benchmark, a benchmark of \benchTotalTasks{} tool-project pairs, together with \analysisagent, a purpose-built agent that achieves \bestVerifiedRate\% verified success.
Our evaluation across four agent architectures and four LLM backends shows that task-specific architecture provides gains that model scaling alone does not, that self-validated success is unreliable without evidence-based validation, and that Java toolchain complexity remains a bottleneck that neither stronger models nor better scaffolding fully resolve.
These results suggest that LLM agents can substantially reduce the effort required to deploy software-analysis tools, but only when their workflows enforce structure and validate concrete evidence. In practice, agent outputs should be validated against concrete artifacts and task-specific criteria before being integrated into CI/CD or developer workflows.

\section{Data Availability}
Our benchmark, agent implementation, and experimental data are publicly available at \url{https://github.com/sola-st/software-analysis-agent}.

\balance

\bibliographystyle{IEEEtran}
\bibliography{references,referencesMichael}
\end{document}

%% file: results-macros.tex
\newcommand{\benchTotalTasks}{35}  
\newcommand{\benchNumTools}{seven}  
\newcommand{\benchNumProjects}{ten}  
\newcommand{\benchCppTools}{four}  
\newcommand{\benchCppProjects}{five}  
\newcommand{\benchJavaTools}{three}  
\newcommand{\benchJavaProjects}{five}  
\newcommand{\benchCppPairs}{20}  
\newcommand{\benchJavaPairs}{15}  

\newcommand{\budgetMaxCycles}{120}  
\newcommand{\budgetCostCap}{2}  
\newcommand{\budgetTimeoutHours}{\SI{5}{\hour}}  
\newcommand{\extendedRunBudget}{\SI{3}{\hour}}  

\newcommand{\bestVerifiedRate}{94}  
\newcommand{\bestVerifiedCount}{33}  
\newcommand{\bestSelfValidatedRate}{100}  

\newcommand{\resAAAvgVerified}{79}  

\newcommand{\resRAGAvgVerified}{10}  

\newcommand{\resSWEAvgVerified}{37}  

\newcommand{\resExecGeminiVerified}{77}  
\newcommand{\resExecAvgVerified}{57}  

\newcommand{\timemNanoMean}{144}  
\newcommand{\timemDeepSeekMean}{47}  
\newcommand{\timemGeminiMeanMin}{36}  

\newcommand{\cycNanoMeanCalls}{78}  
\newcommand{\cycDeepSeekMeanCalls}{56}  
\newcommand{\cycGeminiSuccessCycles}{30}  
\newcommand{\cycConvergenceCycles}{10}  

\newcommand{\ecoJavaMeanMin}{189}  
\newcommand{\ecoCppMeanMin}{110}  
\newcommand{\ecoJavaMeanCalls}{269}  
\newcommand{\ecoCppMeanCalls}{176}  

\newcommand{\effFailCycleMult}{2.77}  
\newcommand{\effFailDurationMult}{4.07}  
\newcommand{\effFailCostMult}{1.27}  
\newcommand{\effFailRetryMult}{1.76}  
\newcommand{\effGeminiFailSensitivity}{3.3}  
\newcommand{\effDeepSeekFailSensitivity}{2.0}  

\newcommand{\failJavaShareOfFailures}{62}  
\newcommand{\failCppShareOfFailures}{38}  

\newcommand{\toolSuccessAFL}{50}  
\newcommand{\toolSuccessCflow}{63}  
\newcommand{\toolSuccessCSA}{54}  
\newcommand{\toolSuccessWALA}{34}  
\newcommand{\toolFailShareWALA}{17.5}  
\newcommand{\toolFailShareInfer}{16.5}  
\newcommand{\toolFailShareSJK}{15.5}  
\newcommand{\toolFailShareKLEE}{15.2}  

\newcommand{\logHeadChars}{1{,}000}  
\newcommand{\logTailChars}{1{,}000}  
\newcommand{\logFailurePatterns}{70}  